\documentclass[conference]{IEEEtran}
\IEEEoverridecommandlockouts
\usepackage[left=0.625in, right=0.625in, top=0.7in, bottom=1in]{geometry}
\usepackage{cite}
\usepackage{amsmath,amssymb,amsfonts}
\usepackage[linesnumbered,ruled,vlined]{algorithm2e}
\SetKwInput{KwInput}{Input}  
\SetKwInput{KwOutput}{Output}
\SetKw{And}{and}
\SetKw{Or}{or}
\SetKw{Break}{break}
\usepackage{algpseudocode}%
\usepackage{graphicx}
\usepackage{textcomp}
\usepackage{xcolor}
\usepackage{subcaption}
\usepackage{multirow}
\usepackage{siunitx}
\usepackage{balance}
\usepackage{lipsum}
\usepackage{mathtools}
\usepackage{cuted}
\usepackage{stfloats}
\usepackage{bbm} % for indicator function
\usepackage{nicefrac}

\usepackage{enumitem}

\usepackage{amsthm}
\usepackage{tabularx}
\makeatletter
\newcommand{\multiline}[1]{%
  \begin{tabularx}{\dimexpr\linewidth-\ALG@thistlm}[t]{@{}X@{}}
    #1
  \end{tabularx}
}
\makeatother

\usepackage[font=small]{caption}

\title{Learning-Based Multiuser Scheduling in MIMO-OFDM Systems with Hybrid Beamforming}
\author{Pouya Agheli$^*$\thanks{*At the time of submission, Pouya Agheli was with the Communication Systems Department at EURECOM, Sophia Antipolis, France. This work was completed during his internship at Nokia Bell Labs.}, Tugce Kobal, Fran\c{c}ois Durand, Matthew Andrews\\Nokia Bell Labs}

\begin{document}

\maketitle
\begin{abstract}
We investigate the multiuser scheduling problem in multiple-input multiple-output (MIMO) systems using orthogonal frequency division multiplexing (OFDM) and hybrid beamforming in which a base station (BS) communicates with multiple users over millimeter wave (mmWave) channels in the downlink. Improved scheduling is critical for enhancing spectral efficiency and the long-term performance of the system from the perspective of proportional fairness (PF) metric in hybrid beamforming systems due to its limited multiplexing gain. Our objective is to maximize PF by properly designing the analog and digital precoders within the hybrid beamforming and selecting the users subject to the number of radio frequency (RF) chains. Leveraging the characteristics of mmWave channels, we apply a two-timescale protocol. On a long timescale, we assign an analog beam to each user. Scheduling the users and designing the digital precoder are done accordingly on a short timescale. To conduct scheduling, we propose combinatorial solutions, such as greedy and sorting algorithms, followed by a machine learning (ML) approach. Our numerical results highlight the trade-off between the performance and complexity of the proposed approaches. Consequently, we show that the choice of approach depends on the specific criteria within a given scenario.
\end{abstract}

\section{Introduction}
Millimeter wave (mmWave) frequency band is a promising candidate for 5G and beyond \cite{rusek2012scaling}. However, mmWave signals are inherently susceptible to high path loss, necessitating base stations (BSs) to be equipped with large antenna arrays and beamforming techniques to mitigate interference and enhance spectral efficiency. \emph{Hybrid beamforming} has gained traction as a viable solution in mmWave systems, offering fewer required radio frequency (RF) chains. Hybrid beamforming enables improved management of multiuser interference and strikes a better balance of performance and efficiency than conventional analog and digital beamforming \cite{alkhateeb2015limited}. Nevertheless, the reduced number of RF chains restricts the multiplexing gain in multiuser systems and limits the maximum number of users that each BS can simultaneously serve. Consequently, \emph{multiuser scheduling} becomes essential such that the BS could dynamically select a subset of users to serve at each time.

The literature contains several studies on hybrid beamforming and user scheduling as part of radio resource management (RRM) for multiple-input multiple-output (MIMO) systems. The hybrid beamforming problem has been explored in \cite{alkhateeb2015limited, alkhateeb2016frequency, park2017dynamic, sohrabi2017hybrid}. In \cite{alkhateeb2015limited}, the authors assume that the number of users is no greater than that of RF chains. Moreover, in \cite{alkhateeb2016frequency, park2017dynamic, sohrabi2017hybrid}, it is assumed that the BS serves a single user, where scheduling is unnecessary. On the other hand, studies such as \cite{kwon2016joint, GomezCZKPBZ22, KimA23, hosseini2022multi, quan2024planning} have investigated the hybrid beamforming and user scheduling as a joint problem under the assumption that the number of users exceeds the available RF chains. That being said, a flat-fading channel model is adopted in \cite{kwon2016joint, GomezCZKPBZ22, KimA23} for mmWave hybrid beamforming systems. In practice, this channel model is less realistic for real-world applications. Addressing this issue, the authors in \cite{hosseini2022multi} have considered a frequency-selective model and formulated the joint user scheduling and hybrid beamforming based on orthogonal frequency division multiplexing (OFDM) systems. As two major limitations, their approach only relies on an offline solution for a relaxed problem, which may be impractical. It also lacks a comprehensive comparison of the performance and complexity across different methods.

This paper falls within the realms of joint hybrid beamforming and user scheduling in MIMO-OFDM systems. We build on the work presented in \cite{KimA23} and extend it to OFDM systems under frequency-selective fading models. We utilize a two-timescale protocol to solve the joint problem of hybrid beamforming and user scheduling. To select a subset of users, we leverage various combinatorial methods along with a machine learning (ML) approach, each offering distinct performance and complexity profiles. Our numerical results reveal that, in terms of long-term proportional fairness and required run time, one of these approaches may be more suitable depending on its delivered trade-off between performance and complexity.

\section{System Model}
\begin{figure}[t!]
    \centering    
    \includegraphics[width=0.43\textwidth, angle=0]{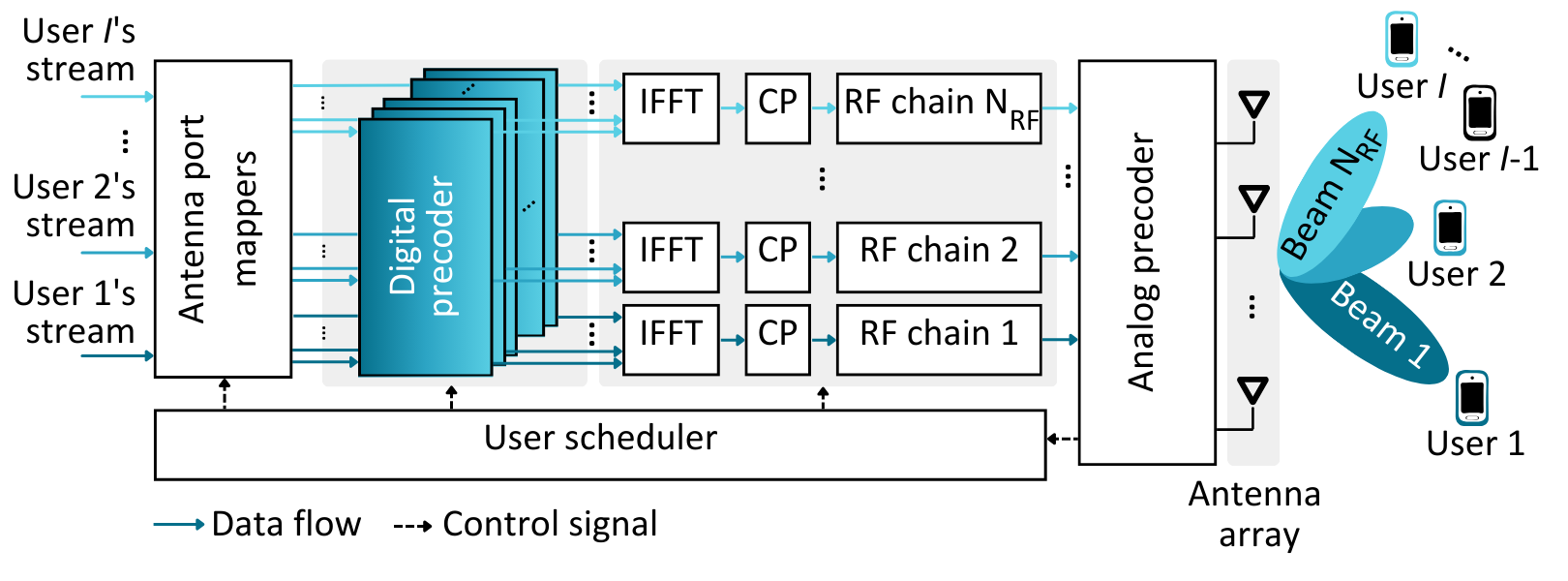}
    \vspace{-0.12cm}
    \caption{A multiuser MIMO-OFDM system in downlink.}
    \label{fig:system_model}
\end{figure}

We consider a downlink multiuser system in which a BS communicates with $I$ \emph{users} in a time-slotted manner over MIMO mmWave channels (Fig.~\ref{fig:system_model}). We assume the channels follow a \emph{frequency-selective} fading model. Therefore, we exert OFDM technology and adopt a hybrid beamforming architecture at the BS assisted by $N_{\rm TX}$ antennas and $N_{\rm RF}$ RF chains, where $N_{\rm RF} < N_{\rm TX}$. In this model, every user has $N_{\rm RX}$ antennas and a single RF chain. Each RF chain at the BS is linked to an independent OFDM \emph{resource grid} that consists of $K$ \emph{physical resource blocks} (PRBs) within a single slot. Every PRB has several subcarriers with equal bandwidths. We assume that \emph{one}\footnote{This assumption distinguishes between OFDM and orthogonal frequency division multiple access (OFDMA). OFDMA allows scheduling multiple users within each resource grid. However, we keep this technology for future studies.} user is scheduled per resource grid at each slot and define $I_{\rm max}$ as the maximum number of users served simultaneously, where $I_{\rm max} \leq I$. Thus, we have $I_{\rm max} = N_{\rm RF}$, and the spatial multiplexing gain of the considered hybrid beamforming system is limited by $\operatorname{min}\{N_{\rm RF}, I\}= N_{\rm RF}$ \cite{alkhateeb2015limited}.
At the time slot $t$, $\forall t\in\mathbb{N}$, a \emph{user scheduler} at the BS selects a subset of users to be served, which is denoted by $\mathcal{M}(t)$ with the size of $\lvert\mathcal{M}(t)\lvert=M(t)$ such that $M(t)\leq I_{\rm max}$, and
\begin{equation}
    \mathcal{M}(t) = \{i_m~|~\forall m=1, 2, ..., M(t)\}.
\end{equation}
% Fig.~\ref{fig:resource_grid} shows the design of $N_{\rm RF}$ resource grids, where $M(t)\leq N_{\rm RF}$ users are scheduled to be served at the $t$-th time slot.
% \begin{figure}[t!]
%     \centering    
%     \includegraphics[width=0.34\textwidth, angle=0]{figures/resource_grid.png}
%     \caption{Design of $N_{\rm RF}$ OFDM resource grids, each having $K$ PRBs.}
%     \label{fig:resource_grid}
% \end{figure}

At the BS, a module with $N_{\rm RF}$ \emph{antenna port mappers} first maps $M(t)$ selected users' data streams to the inputs of $K$ \emph{digital (baseband) precoders}. In this context, an antenna port can be seen as a logical entity associated with an independent resource grid. Thus, each antenna port is uniquely assigned to one of the selected users and maps the modulated symbols of that user's data stream to $K$ PRBs, with each symbol being mapped to \emph{one} distinct subcarrier within a PRB. 
% This is typically done following a specific pattern or order. 
Subsequently, the $k$-th precoder, where $k=1, 2, …, K$, applies baseband precoding to all selected users' symbols arranged side-by-side in the $k$-th PRB.
% At the BS, an \emph{antenna port mapper} first maps selected users' data streams to the inputs of $K$ \emph{digital (baseband) precoders}.  Afterward, the $k$-th precoder, where $k=1, 2, …, K$, applies baseband precoding on $M(t)$ selected users' streams to be transmitted separately through the $k$-th PRB among different resource grids. In this regard, if we consider that the BS can transmit \emph{one} data stream per PRB toward a selected user, 
In this regard, the precoding matrix of the $k$-th baseband precoder at the $t$-th slot is denoted by $\mathbf{F}_{k}(t)$ and defined as follows
\begin{align}\label{eq:bb_precoder}
    \mathbf{F}_{k}(t) = [\mathbf{f}_{k, i_1}(t), ..., \mathbf{f}_{k, i_{M(t)}}(t)] \in \mathbb{C}^{M(t) \times M(t)}.
\end{align}
Afterward, inverse fast Fourier transform (IFFT) and cyclic prefix (CP) adder modules followed by an RF chain at each terminal port dedicated to one selected user construct an OFDM symbol from $K$ digitally precoded symbols of that user. We consider a fully connected structure in which all RF chains are connected to all antennas. Therefore, $M(t)$ OFDM signals are jointly precoded via an \emph{analog precoder} under the precoding matrix $\mathbf{G}(t)$ defined as
\begin{align}\label{eq:ab_precoder}
    \mathbf{G}(t) = [\mathbf{g}_{i_1}(t), ..., \mathbf{g}_{i_{M(t)}}(t)] \in \mathbb{C}^{N_{\rm TX} \times M(t)}.
\end{align}
The analog precoder typically adopts a beamforming \emph{codebook} $\mathcal{G}=\{\mathbf{\psi}_\ell~|~\ell=1, 2, …, \lvert \mathcal{G}\lvert\}$ in mmWave.\footnote{The codebook is defined since phase shifters take quantized angles \cite{alkhateeb2015limited}.} In this scenario, $\mathbf{\psi}_\ell \in \mathbb{C}^{ N_{\rm TX} \times 1}$ represents the $\ell$-th analog beamforming vector. Following the user scheduling, the corresponding beamforming vector for the user $i_m\in\mathcal{M}(t)$ is selected from the codebook, i.e., $\mathbf{g}_{i_m}(t) \in \mathcal{G}$ at the $t$-th slot.

Once the analog beamforming is performed, an outcome signal $\mathbf{x}_k(t) \in \mathbb{C}^{N_{\rm TX} \times 1}$ is transmitted via an antenna array within the $k$-th PRB at the $t$-th slot, which is modeled as
\begin{align}\label{eq:tx_signal}
    \mathbf{x}_k(t) &= \mathbf{G}(t) \mathbf{F}_{k}(t) \mathbf{s}_k(t) \nonumber \\
    &= \mathbf{G}(t) \sum_{i_m \in \mathcal{M}(t)} \mathbf{f}_{k, i_m}(t) s_{k, i_m}(t)
\end{align}
where $\mathbf{s}_k(t) = [s_{k, i_1}(t), ..., s_{k, i_M(t)}(t)]  \in \mathbb{C}^{M(t) \times 1}$ with $s_{k, i_m}(t) \in \mathbb{C}$ being the data stream of the user $i_m \in \mathcal{M}(t)$ to be transmitted within the $k$-th PRB. Here, $\mathbb{E}[\lvert s_{k, i_m}(t) \lvert^2]=1$, and $\mathbb{E}[s^*_{k, i_m}(t)s_{k, j_m}(t)]=0$ for $i_m \neq j_m \in \mathcal{M}(t)$. With regard to \eqref{eq:tx_signal}, the transmission power of the BS is truncated as $\mathbb{E}[\lvert \mathbf{x}_k(t) \lvert^2]\leq P_{k}$ with $P_{k} \in \mathbb{R}^+$ indicating the power limit at the $k$-th PRB. Let us define $\mathbf{h}_{k, i_m}(t) \in \mathbb{C}^{N_{\rm TX} \times N_{\rm RX}}$ as the channel matrix between the BS and the $i_m$-th user through the $k$-th PRB at the $t$-th slot. The final processed signal $y_{k, i_m}(t) \in \mathbb{C}$ at that user is derived as
\begin{align}\label{eq:rx_signal}
    y_{k, i_m}(t) = \widehat{\mathbf{g}}_{i_m}^H(t)\mathbf{h}^H_{k, i_m}(t) \mathbf{x}_k(t) + \widehat{\mathbf{g}}_{i_m}^H(t)\mathbf{n}_{i_m}(t)
\end{align}
where $\mathbf{n}_{i_m}(t) \sim \mathcal{CN}(0,\mathbf{\sigma}_{i_m}^2) \in \mathbb{C}^{N_{\rm RX} \times 1}$ is a white Gaussian noise, and $\mathbf{\sigma}_{i_m}^2$ shows the noise power. 
In addition, $\widehat{\mathbf{g}}_{i_m}(t) \in \mathbb{C}^{N_{\rm RX} \times 1}$ denotes the analog combining vector that an \emph{analog combiner} at the user derives and then applies to the received signal. Importing \eqref{eq:rx_signal} into \eqref{eq:tx_signal}, the achievable rate for the user $i_m\in \mathcal{M}(t)$ is computed as
\begin{align}\label{eq:rate_user}
    r_{i_m}(t) = \sum_{k=1}^{K} B_k \log_2\!\left(1 + {\rm SINR}_{k, i_m}\right)~~[\text{bits}/\text{sec}/\text{Hz}]
\end{align}
at the $t$-th slot, whereas $r_{j_m}(t)=0$, $\forall j_m \notin \mathcal{M}(t)$. In \eqref{eq:rate_user}, $B_k$ denotes the \emph{bandwidth factor} of the $k$-th PRB, where its value is equal to the product of the number of subcarriers per PRB and the number of OFDM symbols per slot. Also, we have
\begin{align}\label{eq:sinr}
     {\rm SINR}_{k, i_m} =\frac{\lvert \mathbf{u}_{k, i_m}(t) \mathbf{f}_{k, i_m}(t)\lvert^2}{\sum\limits_{j_m \in \mathcal{M}(t)\setminus\{i_m\}} \lvert \mathbf{u}_{k, i_m}(t) \mathbf{f}_{k, j_m}(t)\lvert^2 + \mathbf{\sigma}_{i_m}^2}
\end{align}
where $\mathbf{u}_{k, i_m}(t) = [u_{k,i_mj_1}, ..., u_{k,i_mj_m}]^H \in \mathbb{C}^{M(t) \times 1}$ with its element $u_{k,i_mj_m} = \widehat{\mathbf{g}}_{i_m}^H(t)\mathbf{h}^H_{k, i_m}(t) \mathbf{g}_{j_m}(t)\in\mathbb{C}$, $\forall i_m,j_m \in \mathcal{M}(t)$, which is called an \emph{effective channel}.

% Based on \eqref{eq:rate_user} and \eqref{eq:sinr}, the maximization problem is defined in the subsequent section.

\section{Proportional Fairness Maximization}
Within this section, we first define a \emph{proportional fairness} (PF) metric and then formulate the optimization problem.

\subsection{The PF Metric}
Let us consider $R_{i}(T) = (1-\eta_i) R_{i}(T-1) + \eta_i r_{i}(T)$ the cumulative data rate of the $i$-th user within the period of $T$ time slots in the form of an exponential moving average of the rates \cite{andrews2007survey}, where $r_{i}(t)$ denotes the instantaneous rate from \eqref{eq:rate_user}, and $0\leq\eta_i\leq 1$ is a tuning factor. To initialize, we have $R_{i}(0)=1$ for $i=1, 2, ..., I$. The PF metric is defined as
\begin{align}\label{eq:pf_metric}
    \operatorname{PF} = \sum_{i=1}^{I} \log(R_{i}(T)),
\end{align}
which guarantees that \emph{no} user is starved completely. 
% Based on \eqref{eq:pf_metric}, the optimization problem is formulated in Section~\ref{sec:3-b}.

\subsection{The Optimization Problem}\label{sec:3-b}
The long-term objective is to maximize the defined $\operatorname{PF}$ in \eqref{eq:pf_metric}, subject to the constraints imposed by the hybrid beamforming system. Inspired from \cite{stolyar2005maximizing}, maximizing PF within the period of $T$ slots could be decomposed into maximizing the weighted sum of instantaneous data rates, i.e., $\sum_{i=1}^{I} w_i(t) r_i(t)$ at each slot $t=1, 2, ..., T$, as consecutive \emph{one-shot} problems. In this context, we consider $w_i(t) = 1 / R_i(t-1)$, which offers user fairness such that the user scheduler would select the users that have not been served for a long time. Therefore, the one-shot problem $\mathcal{P}$ at the $t$-th slot is formulated as follows
\begin{align}\label{eq:opt_max}
    &\mathcal{P} :  ~\underset{\mathcal{M}(t),\, \{\mathbf{F}_{k}(t)\}_{k=1}^{K},\, \mathbf{G}(t)}{\operatorname{max}} ~\sum_{i=1}^{I} w_i(t)r_i(t) \nonumber \\
    & {\rm s.t.}~ 
    \mathcal{C}_1: \mathcal{M}(t) \subset \{1, 2, ..., I\}, ~~ \mathcal{C}_2: \lvert\mathcal{M}(t)\lvert\leq I_{\rm max}, \nonumber  \\
    &~~~~~ \mathcal{C}_3: \sum_{i_m\in\mathcal{M}(t)}  \left\lVert \mathbf{G}(t)\mathbf{f}_{k, i_m}(t) \right\lVert^2 \leq P_{k},\, k=1,..., K\nonumber  \\
    &~~~~~ \mathcal{C}_4: \mathbf{g}_{i_m}(t) \in \mathcal{G},\, \forall i_m\in \mathcal{M}(t),
\end{align}
where $\mathcal{C}_1$ and $\mathcal{C}_2$ are user scheduling requirements, the transmission power constraint at the BS is shown in $\mathcal{C}_3$, and $\mathcal{C}_4$ implies the codebook utilization for analog RF beamforming.

The problem $\mathcal{P}$ is a mixed-integer problem (MIP) as $\mathcal{M}(t)$ and $\mathbf{G}(t)$ exist in discrete spaces, whereas $\mathbf{F}_{k}(t)$, for $k=1, 2, ..., K$, has a continuous outcome space. The total number of possible combinations for the discrete solution variables is derived as 
$\sum_{i=1}^{I_{\rm max}} \big(\!\binom{I}{i} \times \lvert\mathcal{G}\lvert^i\big)$, which grows fast with the increase of $I_{\rm max}$, $I$, and $\lvert\mathcal{G}\lvert$. Also, to compute the users' data rates, i.e., $r_i(t)$ from \eqref{eq:rate_user}, the BS needs to know the channel information of all users within every PRB in advance. This requires the users to estimate their channels and share them with the BS, resulting in a large time overhead and waste of energy. To address this, we develop a \emph{two-timescale} protocol.

\section{Tailored Two-Timescale Protocol}\label{sec:4}
% In this section, we present the two-timescale protocol and then reformulate the problem $\mathcal{P}$ based on this protocol.

\subsection{Two-Timescale Protocol}\label{sec:4-a}
\begin{figure}[t!]
    \centering    
    \includegraphics[width=0.287\textwidth, angle=0]{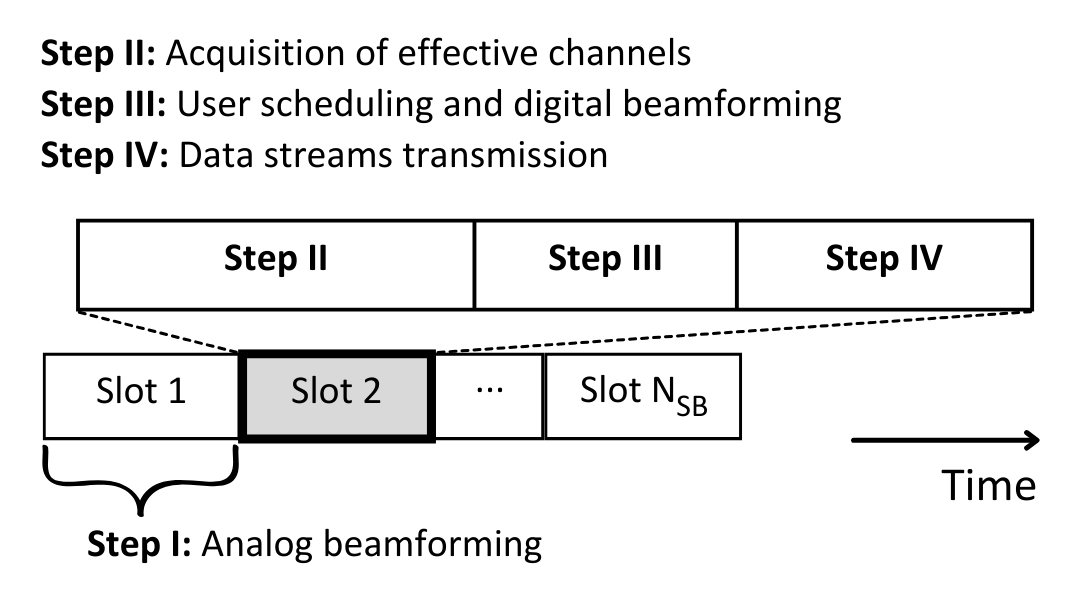}
    \caption{Structure of the two-timescale protocol for one resource grid.}
    \label{fig:protocol}
\end{figure}

The two-timescale protocol exploits mmWave characteristics, where path gains vary slower than path angles \cite{akdeniz2014millimeter, va2016beam, cai2020two}. We assume fixed path angles over long-time blocks, while correlated gains may change among shorter blocks. Given the limited number of clusters with distinct path angles in every mmWave channel, relative to $N_{\rm TX}$, we can apply \emph{directional} analog beamforming on a \emph{long} timescale such that the highest gain of the beamformer is aligned with the dominant path angle \cite{akdeniz2014millimeter}. Digital beamforming adapts to path gains on a \emph{short} timescale, along with user scheduling at each short-time block. The two-timescale protocol is shown in Fig.~\ref{fig:protocol}, where a long-time block consists of $N_{\rm SB}$ short-time blocks for a resource grid, indexed by $t\in\mathbb{N}$. The detailed steps follow.

\textbf{Step~I. \textit{Analog beamforming:}} During the first short-time block of each long-time block, i.e., $t=1, N_{\rm SB}+1, ...$, the BS broadcasts a pilot symbol to all users by applying each beamforming vector $\mathbf{\psi}_\ell$ in the codebook $\mathcal{G}$ sequentially for $\ell=1, 2, …, \lvert \mathcal{G}\lvert$. Then, each user selects its best beam based on the received pilots and its derived analog combining vector. The index of the best beam for the $i$-th user is derived as 
\begin{align}
    \ell_i^*(t) = ~\underset{\ell=1, 2, …, \lvert \mathcal{G}\lvert}{\operatorname{arg\,max}} \frac{1}{K}\sum_{k=1}^{K} \left|\widehat{\mathbf{g}}_{i}^H(t) \mathbf{h}^H_{k, i}(t)\mathbf{\psi}_\ell \right|^2.
\end{align}
Afterward, each user shares the index of its best beam with the BS, and the BS assigns an analog beamforming vector from the codebook $\mathcal{G}$ to each user accordingly so that the analog precoder of the $i$-th user is derived as $\mathbf{g}_{i}^*(t) = \mathbf{\psi}_{\ell_i^*(t)}$. From \eqref{eq:ab_precoder}, $\mathbf{G}^*(t)$ is constructed at time slots $t=1, N_{\rm SB}+1, ...$ and fixedly used through the following $N_{\rm SB}-1$ short-time blocks.

\textbf{Step~II. \textit{Acquisition of effective channels:}} At the $t$-th short-time block, the BS generates an effective channel matrix denoted by $\mathbf{U}_{k}(I;t)=[\mathbf{u}^*_{k, 1}(t),...,\mathbf{u}^*_{k, I}(t)] \in \mathbb{C}^{I \times I}$. For this, the $i$-th user measures the effective channels $u^*_{k,ij}=\widehat{\mathbf{g}}_{i}^H(t)\mathbf{h}^H_{k, i}(t) \mathbf{g}^*_{j}(t)\in\mathbb{C}$ for $i,j=1,...,I$ within the $k$-th PRB based on the assigned beams and then shares $\mathbf{u}^*_{k, i}(t) = [u^*_{k,i1}, ..., u^*_{k,iI}]^H \in \mathbb{C}^{I \times 1}$ for $k=1,...,K$ with the BS. 

\textbf{Step~III. \textit{User scheduling and digital beamforming:}} According to the assigned analog beamforming vectors, i.e., $\mathbf{G}^*(t)$, the generated effective channel matrix, i.e., $\mathbf{U}_{k}(I;t)$, and weights of the users, i.e., $\{w_i(t)\}_{i=1}^{I}$, from their cumulative data rates, the BS conducts user scheduling and digital beamforming at the $t$-th short-time block. Hence, $\mathcal{P}$ can be reformed to reduce complexity, as discussed in Section~\ref{sec:4-b}.

\textbf{Step~IV. \textit{Data transmission:}} With the designed system, the selected user $i_m$, $\forall i_m \in \mathcal{M}(t)$, at the $t$-th short-time block is served under the rate of $r_{i_m}(t)$ as in \eqref{eq:rate_user}. The weights of all users are updated based on their cumulative data rates.

Steps II to IV are repeated for $N_{\rm SB}-1$ short-time blocks starting from the second block until the end of each long-time block, i.e., $t=2,..., N_{\rm SB}$ for the first long-time block. 
% The same procedure holds for the other long-time blocks.
    
\subsection{Problem of User Scheduling and Digital Beamforming}\label{sec:4-b}
We reach a new problem with reduced complexity owing to the two-timescale protocol from Section~\ref{sec:4-a}. 
% In this case, the protocol and the users' feedback allow for assigning the best analog beam to each user. 
After selecting users, i.e., $\mathcal{M}(t)$, the precoding matrix is derived as 
\begin{align}\label{eq:ab_precoder_new}
    \mathbf{G}^*(t) = [\mathbf{g}^*_{i_1}(t), ..., \mathbf{g}^*_{i_{M(t)}}(t)].
\end{align}
Having $\mathbf{G}^*(t)$, $\mathcal{P}$ in \eqref{eq:opt_max} is transformed into a joint problem of user scheduling and digital beamforming as follows
\begin{align}\label{eq:opt_max_reformed}
    &\widehat{\mathcal{P}} :  ~\underset{\mathcal{M}(t),\, \{\mathbf{F}_{k}(t)\}_{k=1}^{K}}{\operatorname{max}} ~\sum_{i=1}^{I} w_i(t)r_i(t) \nonumber \\
    & {\rm s.t.}~ 
    \mathcal{C}_1: \mathcal{M}(t) \subset \{1, 2, ..., I\}, ~~ \mathcal{C}_2: \lvert\mathcal{M}(t)\lvert\leq I_{\rm max}, \nonumber  \\
    &~~~~~ \mathcal{C}_3: \sum_{i_m\in\mathcal{M}(t)}  \left\lVert \mathbf{G}^*(t)\mathbf{f}_{k, i_m}(t) \right\lVert^2 \leq P_{k},\, k=1,..., K.
\end{align}
With a given $\mathcal{M}(t)$ at the $t$-th time slot, the precoding matrix of the $k$-th digital precoder, i.e., $\mathbf{F}_{k}(t)$, is derived by solving $\widehat{\mathcal{P}}$ in \eqref{eq:opt_max_reformed}, with regard to \eqref{eq:rate_user} and \eqref{eq:sinr}. 

To design the digital precoder, we apply \emph{zero-forcing} (ZF) beamforming \cite{yoo2006optimality}, which offers near-optimal performance in mmWave due to the small number of clusters \cite{alkhateeb2015limited, akdeniz2014millimeter}. Based on $\mathcal{M}(t)$ in hand at the $t$-th short block, the effective channel matrix is derived concerning the selected users, as below
\begin{align}\label{eq:eff_matrix}
    \mathbf{U}_{k}(\mathcal{M}(t);t)=[\mathbf{u}^*_{k, i_1}(t),...,\mathbf{u}^*_{k, i_m}(t)] \in \mathbb{C}^{{M}(t) \times {M}(t)}.
\end{align}
From \eqref{eq:eff_matrix}, the ZF-based digital precoder for the $k$-th PRB satisfies $\mathbf{U}_{k}(\mathcal{M}(t);t)\mathbf{F}_{k}(t)=\mathbf{I}_{M(t)}$, where $\mathbf{I}_{M(t)}$ denotes the identity matrix of size $M(t)$. Thus, we can write
\begin{align}
    \mathbf{F}_{k}(t) = \mathbf{U}^H_{k}(\mathcal{M}(t);t)\Big[\mathbf{U}_{k}(\mathcal{M}(t);t)\mathbf{U}^H_{k}(\mathcal{M}(t);t)\Big]^{-1}.
\end{align}
Considering a fair power allocation scheme for all $M(t)$ data streams within each PRB, we have $\left\lVert \mathbf{G}^*(t)\mathbf{f}_{k, i_m}(t) \right\lVert^2 = \frac{P_{k}}{M(t)}$ for $k=1,...,K$. Hence, the beamforming vector for the user $i_m$, $\forall i_m \in \mathcal{M}(t)$, in $\mathbf{F}_{k}(t)$ is normalized such that
\begin{align}\label{eq:db_new}
    \mathbf{f}_{k, i_m}(t) = \sqrt{\frac{P_{k}}{M(t)}} \frac{\mathbf{f}_{k, i_m}(t)}{\left\lVert \mathbf{G}^*(t)\mathbf{f}_{k, i_m}(t) \right\lVert_2}.
\end{align}
By the use of \eqref{eq:ab_precoder_new}, \eqref{eq:eff_matrix}--\eqref{eq:db_new}, and the given $\mathcal{M}(t)$, we compute $r_{i_m}(t)$ for the user $i_m$, $\forall i_m \in \mathcal{M}(t)$, from \eqref{eq:rate_user} and \eqref{eq:sinr}. Accordingly, we can define $\bar{r}(\mathcal{M}(t);t)=\sum_{i_m \in \mathcal{M}(t)}w_{i_m}(t)r_{i_m}(t)$ as the \emph{weighted sum-rate} of the scheduled users at the $t$-th block.

To conduct scheduling, we aim to find the best set of users at each short block to maximize the sum-rate at the respective slot. In this sense, a naive approach is to employ the \emph{brute-force} algorithm or perform an \emph{exhaustive search} by calculating the ZF-based digital precoder for every possible combination of $M(t)$ users and then selecting the best case that yields the highest $\bar{r}(\mathcal{M}(t);t)$. However, either algorithm demands heavy computation when dealing with large $I$ and $M(t)$. 
% Therefore, in Sections~\ref{sec:5} and \ref{sec:6}, we delve into alternative user scheduling approaches with potentially lower computational overhead.

\section{Multiuser Scheduling Design Using Combinatorial Approaches}\label{sec:5}
% Through this section, we explore combinatorial scheduling approaches in the form of \emph{greedy} and \emph{sorting} algorithms.

\subsection{Greedy Algorithm}\label{sec:5-a}
% We can demonstrate the greedy algorithm from the perspectives of \emph{incremental} and \emph{decremental} approaches, as follows.
\subsubsection{Incremental approach}
\begin{algorithm}[t!]
    \caption{Greedy incremental user scheduling} \label{alg:alg1}
    \KwInput{$I$, $I_{\rm max}$, $\{\mathbf{u}_{k, i}(t)\}_{i=1}^{I}$, and $\{\mathbf{f}_{k, i}(t)\}_{i=1}^{I}$ for $k=1, 2, ..., K$.
    }
    \KwOutput{The set of the scheduled users, i.e., $\mathcal{M}(t) \subset \{i~|~i=1, 2, ..., I\}$.}
    Initialize $\mathcal{M}(t)=\varnothing$ and $\bar{r}(\mathcal{M}(t);t)=0$.\\
    \For{$n=1, 2, ..., I_{\rm max}$}{
    Find the $i^*$-th user that offers the highest weighted sum-rate user by the use of \eqref{eq:greedy_inc}.\\
    \lIf{$\bar{r}(\mathcal{M}(t) \cup \{i^*\};t) > \bar{r}(\mathcal{M}(t);t)$}{$\mathcal{M}(t) \gets \mathcal{M}(t) \cup \{i^*\}$}
    \lElse{\Break}
    }
    \KwRet $\mathcal{M}(t)$.
\end{algorithm}
The idea behind the incremental algorithm, as illustrated in Algorithm~\ref{alg:alg1}, is to start from an empty set $\mathcal{M}(t)=\varnothing$ at the $t$-th short-time block and add users iteratively until the performance can no longer be improved. Within the $n$-th iteration, where $n=1, 2, ..., I_{\rm max}$, the $i^*$-th user that offers the highest weighted sum-rate is found by 
\begin{align}\label{eq:greedy_inc}
    i^*=\underset{i\in\{i|i=1, 2,..., I\} \setminus \mathcal{M}(t)}{\operatorname{arg\,max}} \bar{r}(\mathcal{M}(t) \cup \{i\};t).
\end{align}
The user $i^*$ is added to $\mathcal{M}(t)$ if it improves the performance, i.e., $\bar{r}(\mathcal{M}(t) \cup \{i^*\};t) > \bar{r}(\mathcal{M}(t);t)$. The process terminates if $\bar{r}(\mathcal{M}(t) \cup \{i^*\};t) \leq \bar{r}(\mathcal{M}(t);t)$ or $M(t)$ reaches $I_{\rm max}$.

\subsubsection{Decremental approach}
% \begin{algorithm}[t!]
%     \caption{Greedy decremental user scheduling} \label{alg:alg2}
%     \KwInput{$I$, $I_{\rm max}$, $\{\mathbf{u}_{k, i}(t)\}_{i=1}^{I}$, and $\{\mathbf{f}_{k, i}(t)\}_{i=1}^{I}$ for $k=1, 2, ..., K$.
%     }
%     \KwOutput{The set of the scheduled users, i.e., $\mathcal{M}(t) \subset \{i~|~i=1, 2, ..., I\}$.}
%     Set $\mathcal{M}(t)=\{i~|~i=1, 2, ..., I\}$ and $\bar{r}(\mathcal{M}(t);t)=0$.\\
%     \For{$n=1, 2, ..., I$}{
%     Derive the $j$-th user that satisfies \eqref{eq:greedy_dec}.\\
%     \lIf{$\bar{r}(\mathcal{M}(t) \setminus \{j\};t) > \bar{r}(\mathcal{M}(t);t)$ \Or $M(t) > I_{\rm max}$}{$\mathcal{M}(t) \gets \mathcal{M}(t) \setminus \{j\}$}
%     \lElse{\Break}
%     }
%     \KwRet $\mathcal{M}(t)$.
% \end{algorithm}
The greedy decremental algorithm starts by initiating a full set of all users, i.e., $\mathcal{M}(t)=\{i~|~i=1, 2, ..., I\}$, from a contrary perspective compared to the incremental algorithm. In the  $n$-th arbitrary iteration, the $j$-th user that satisfies 
\begin{align}\label{eq:greedy_dec}
    j=\underset{j \in \mathcal{M}(t)}{\operatorname{arg\,max}}~ \bar{r}(\mathcal{M}(t) \setminus \{j\};t)
\end{align}
is removed from $\mathcal{M}(t)$. The procedure continues down to the subset of users that ensures the highest weighted sum-rate, i.e., $\bar{r}(\mathcal{M}(t) \setminus \{j\};t) \leq \bar{r}(\mathcal{M}(t);t)$, $\forall j\in \mathcal{M}(t)$, and $M(t) \leq I_{\rm max}$. 
% Algorithm~\ref{alg:alg2} depicts the decremental approach.

The greedy incremental and decremental algorithms require $M(t)(2I - M(t) + 1) / 2$ and $(I-M(t))(I+M(t)+1)/2$ number of user searches to schedule $M(t)$ users, respectively \cite{sun2009mmse}. Thus, an increase in the number of users monotonically increases the search complexity. Additionally, recalculating ZF-based digital precoders within each iteration imposes heavy computational complexity in both approaches.
% This motivates us to explore the sorting algorithm, as comes in the subsequent part.

\subsection{Sorting Algorithm}
The sorting algorithm relies on achievable rates of all users, assuming that the users do not cause any interference with each other. This allows us to obtain a diagonal effective channel matrix with $u_{k,ij}=0$, $\forall i\neq j$, for the $k$-th PRB and design the digital precoder. Then, we can derive rates, sort them in descending order, and pick $M(t)$ users with the highest rates. 
% The value of $M(t)$ is given or can be determined adaptively by running the algorithm for $M(t)=1, 2, ..., I_{\rm max}$ and choosing the best $M(t)$. 
% This algorithm is computationally efficient as the ZF-based precoding is performed once in each block. 
% However, its performance may decrease with an increase in the number of users, considering that the interference factor is ignored.
% We take advantage of a learning-based approach to play with the trade-off between the computational complexity and the performance, as discussed in Section~\ref{sec:6}.

\section{Learning-Based Multiuser Scheduling}\label{sec:6}
% Within this section, we discuss the ML-based approach for multiuser scheduling, detailing our neural network architecture and the sample preparation process for training and evaluation.

\subsection{Neural Network Setup}
To design learning-based scheduling, we employ supervised learning to train a fully connected \emph{deep neural network} (DNN)  with $L_0$ \emph{input features} and $I$ outputs. The DNN consists of \emph{three} hidden layers, where the first, second, and third layers have $L_1$, $L_2$, and $L_3$ nodes, respectively. A \emph{sigmoid} activation function is utilized among all layers. 
% In the following parts, we discuss the input and output setups of the proposed DNN.

\subsubsection{Input features}
The reasonable set of the input features must encapsulate a combination of the designed analog precoder, i.e., $\mathbf{G}^*(t)$, in Step~I of the protocol, the derived effective channel matrix, i.e., $\mathbf{U}_{k}(I;t)$, in Step~II, and the known weights of the users, i.e., $\{w_i(t)\}_{i=1}^{I}$, from Step~IV. Our model takes an input set of features in the form of a vector of size $2I^2 + I(N_{\rm TX}+1)$, which is comprised of \emph{four} subvectors. The first and the second subvectors contain the amplitudes and angles of the effective channel averaged over all PRBs, i.e., $\sum_{k=1}^{K}u_{k,ij} / K$, for $i,j=1, 2, ..., I$, respectively, each with the size of $I^2$. The third subvector encompasses the angles of the analog beams assigned to the users, having the size of $N_{\rm TX} I$. The fourth subvector carries the users' weights with the size of $I$. All subvectors are \emph{normalized} to tackle different orders of magnitude between different subvectors. 
% After investigating several combinations, we reached this set of features that balances complexity and performance. Specifically, we attempted to remove each subvector one by one, add other subvectors, such as the one containing the amplitudes of the assigned analog beams, and change the order of the chosen subvectors. After analyzing the offered performance and induced complexity of all combinations, we landed on the set of input features that offers the best trade-off.

\subsubsection{Output decisions}
Multiuser scheduling can be seen as a \emph{binary classification problem} where the users are classified as \emph{selected} or \emph{deselected}. We consider that the $i$-th \emph{rounded} output of the DNN is associated with the selection or deselection of the $i$-th user, where $i=1, 2, ..., I$. Let us define $\alpha_i$ as a selection variable for the $i$-th user, where $\alpha_i=1$ if that user is selected; otherwise, $\alpha_i=0$. Thus, the set of the scheduled users according to the outputs of the DNN is derived by
\begin{align}
    \widehat{\mathcal{M}}(t)=\{i~|~\alpha_i=1, \forall i=1, 2, ...,I\}.
\end{align}
As the DNN could potentially make an incorrect decision that results in $\lvert\widehat{\mathcal{M}}(t)\lvert > I_{\rm max}$, we cascade a selection filter with the DNN. The filter sorts the users in $\widehat{\mathcal{M}}(t)$ based on their weights and selects the top $I_{\rm max}$ of them, which constructs the corresponding $\mathcal{M}(t)$. If $\lvert\widehat{\mathcal{M}}(t)\lvert \leq I_{\rm max}$, inputs are directly forwarded to the output such that $\mathcal{M}(t)=\widehat{\mathcal{M}}(t)$.

\subsection{Training and Evaluation}
A sample set is generated using greedy incremental user scheduling over $N_1$ episodes, each spanning $N_2$ slots. Every episode involves varying initial channel realizations, user distributions, and movement directions. It is considered to allocate a fraction $0\leq\beta\leq1$ of the generated samples for training the model while reserving the remaining $1-\beta$ for evaluating the trained model. 
% These sets are referred to as the \emph{training} and \emph{evaluation} sets, respectively. 
We break the training set into batches of size $N_3$ and train the model for $N_4$ epochs. We use the \emph{cross-entropy loss} function to evaluate performance.

\section{Numerical Results}
% In this section, we evaluate the performance of the proposed hybrid beamforming and user scheduling approaches.
% from various perspectives, such as PF, weighted sum-rate, and run time required by different scheduling methods.

\subsection{Setup and Assumptions}\label{sec:7-a}
We assume $I=20$ users, each with a single antenna, i.e., $N_{\rm RX}=1$, are randomly distributed around the BS in a circular area of $100\,[\text{m}]$ radius. The BS is assumed to be located at a height of $7\,[\text{m}]$. Its antenna array has $N_{\rm TX}=16$ antennas arranged in a \emph{uniform planar array} (UPA) structure, with $8$ elements in the horizontal and $2$ in the vertical. The antenna boresight of the BS is tilted downward by $10^\circ$, where the BS can cover horizontally from $-180^\circ$ to $180^\circ$ and vertically from $-30^\circ$ to $30^\circ$. The analog precoding codebook, i.e., $\mathcal{G}$, consists of a $32\times 8$ grid of beams evenly spaced in the horizontal and vertical directions. Regarding the OFDM technology, we adopt numerology $\mu=5$ from 5G NR \cite{3gpp.38.211}, where a subframe consists of $32$ slots.
Every slot contains $14$ OFDM symbols, each having a duration of $2.23\,[\mu\text{sec}]$ including a CP of $0.15\,[\mu\text{sec}]$. Also, $12$ subcarriers build one PRB. This results in a \emph{subcarrier spacing} (SCS) of $480\,[\text{KHz}]$ and bandwidth factor $B_k=168$ for $k=1, 2, ..., K$. Unless otherwise specified, we use the remaining parameter values listed in Table~\ref{tab:sim_params}.
\noindent
\renewcommand{\arraystretch}{1}
\begin{table}[!t]
\begin{center}
\caption{Parameters for Numerical Results}\label{tab:sim_params}
\begin{tabular}{| l | c | c |}
\hline
\rule{0pt}{8pt}\footnotesize \!\textbf{Name} & \footnotesize \!\textbf{Symbol}\! &\footnotesize \!\textbf{Value}\!\\
\hline
% \hline
% \rule{0pt}{7pt}\footnotesize \!No. users&\footnotesize \!$I$\!&\footnotesize \!$20$\!\\
% \hline
% \rule{0pt}{7pt}\footnotesize \!No. antennas at the BS&\footnotesize \!$N_{\rm TX}$\!&\footnotesize \!$16$\!\\
% \hline
% \rule{0pt}{7pt}\footnotesize \!No. antennas at each user&\footnotesize \!$N_{\rm RX}$\!&\footnotesize \!$1$\!\\
\hline
\rule{0pt}{7pt}\footnotesize \!No. RF chains at the BS&\footnotesize \!$N_{\rm RF}$\!&\footnotesize \!$8$\!\\
\hline
% \rule{0pt}{7pt}\footnotesize \!No. subcarriers per PRB&\footnotesize \!$-$\!&\footnotesize \!$12$\!\\
% \hline
\rule{0pt}{7pt}\footnotesize \!Maximum no. served users&\footnotesize \!$I_{\rm max}$\!&\footnotesize \!$8$\!\\
\hline
\rule{0pt}{7pt}\footnotesize \!Power limit at the $k$-th PRB&\footnotesize \!$P_k$\!&\footnotesize \!\!$20\,[\text{dBm}]$\!\!\\
\hline
\rule{0pt}{7pt}\footnotesize \!Noise power at the $i_m$ user&\footnotesize \!$\mathbf{\sigma}_{i_m}^2$\!&\footnotesize \!\!$-30\,[\text{dBm}]$\!\!\\
\hline
\rule{0pt}{7pt}\footnotesize \!Carrier frequency&\footnotesize \!$f_c$\!&\footnotesize \!\!$28\,[\text{GHz}]$\!\!\\
% \hline
% \rule{0pt}{7pt}\footnotesize \!OFDM numerology&\footnotesize \!$-$\!&\footnotesize \!$1$\!\\
% \hline
\hline
\rule{0pt}{7pt}\footnotesize \!No. PRBs&\footnotesize \!$K$\!&\footnotesize \!$12$\!\\
\hline
\rule{0pt}{7pt}\footnotesize \!No. OFDM subframes per frame&\footnotesize \!$-$\!&\footnotesize \!$10$\!\\
\hline
\rule{0pt}{7pt}\footnotesize \!Duration of each OFDM subframe&\footnotesize \!$-$\!&\footnotesize \!$10^{-3}\,[\text{sec}]$\!\\
\hline
% \rule{0pt}{7pt}\footnotesize \!No. OFDM slots per subframe&\footnotesize \!$-$\!&\footnotesize \!$32$\!\\
% \hline
% \rule{0pt}{7pt}\footnotesize \!No. OFDM subcarriers per PRB&\footnotesize \!$-$\!&\footnotesize \!$12$\!\\
% \hline
% \rule{0pt}{7pt}\footnotesize \!No. OFDM symbols per slot&\footnotesize \!$-$\!&\footnotesize \!$14$\!\\
% \hline
% \rule{0pt}{7pt}\footnotesize \!Duration of each OFDM symbol with CP\!&\footnotesize \!$-$\!&\footnotesize \!\!$2.23\,[\mu\text{sec}]$\!\!\\
% \hline
% \rule{0pt}{7pt}\footnotesize \!OFDM subcarrier spacing (SCS)&\footnotesize \!$-$\!&\footnotesize \!\!$480\,[\text{KHz}]$\!\!\\
% \hline
% \rule{0pt}{7pt}\footnotesize \!Bandwidth factor of the $k$-th PRB&\footnotesize \!$B_k$\!&\footnotesize \!$168$\!\\
% \hline
\rule{0pt}{7pt}\footnotesize \!No. long-time blocks&\footnotesize $-$&\footnotesize \!$100$\!\\
\hline
\rule{0pt}{7pt}\footnotesize \!No. slots per long-time block&\footnotesize \!$N_{\rm SB}$\!&\footnotesize \!$1$\!\\
\hline
\rule{0pt}{7pt}\footnotesize \!No. input features in the DNN&\footnotesize \!$L_0$\!&\footnotesize \!$1140$\!\\
\hline
\rule{0pt}{7pt}\footnotesize \!No. nodes in the DNN's hidden layers\!&\footnotesize \!\!$(L_1, L_2, L_3)$\!\!&\footnotesize \!\!$(1200, 500, 200)$\!\!\\
\hline
% \rule{0pt}{7pt}\footnotesize \!No. nodes in the DNN's second hidden layer&\footnotesize \!$L_2$\!&\footnotesize \!$500$\!\\
% \hline
% \rule{0pt}{7pt}\footnotesize \!No. nodes in the DNN's third hidden layer&\footnotesize \!$L_3$\!&\footnotesize \!$200$\!\\
% \hline
\rule{0pt}{7pt}\footnotesize \!No. sampling episodes&\footnotesize \!$N_1$\!&\footnotesize \!$120$\!\\
\hline
\rule{0pt}{7pt}\footnotesize \!No. slots per sampling episode&\footnotesize \!$N_2$\!&\footnotesize \!$100$\!\\
\hline
\rule{0pt}{7pt}\footnotesize \!Sample splitting fraction&\footnotesize \!$\beta$\!&\footnotesize \!$0.8$\!\\
\hline
\rule{0pt}{7pt}\footnotesize \!Length of each DNN training batch&\footnotesize \!$N_3$\!&\footnotesize \!$16$\!\\
\hline
\rule{0pt}{7pt}\footnotesize \!No. DNN training epochs&\footnotesize \!$N_4$\!&\footnotesize \!$300$\!\\
\hline
\end{tabular}
\medskip
\end{center}
\end{table}

% \begin{figure}[t!]
%     \centering    
%     \includegraphics[width=0.36\textwidth, angle=0]{figures/model_loss.png}
%     \caption{Loss comparison of the scikit-learn and PyTorch libraries.}
%     \label{fig:loss}
% \end{figure}
% Assessing the well-known scikit-learn and PyTorch libraries for learning-based user scheduling, we adopt the latter due to its superior performance in minimizing model loss.

\subsection{Channel Model}
We use the mmWave channel model from \cite[Eq.~(9)]{akdeniz2014millimeter} to derive the channel matrix, i.e., $\mathbf{h}_{k, i}(t)$, between the BS and the $i$-th user within the $k$-th PRB at the $t$-th. Accordingly, we consider the same large-scale parameters as given in \cite{akdeniz2014millimeter} for a carrier frequency of $f_{c} =28\,[\text{GHz}]$. However, to model the small-scale fading, we assume $20$ subpaths within each cluster and a maximum \emph{Doppler shift} of $f_{c,\rm max}=258\,[\text{Hz}]$ concerning the carrier frequency. Since the Doppler shift is a function of frequency, its value changes among different PRBs, each having a bandwidth of $12 \times 480\,[\text{KHz}]$. In this sense, for modeling $\mathbf{h}_{k, i}(t)$, the induced Doppler shift $f^\prime_{k,\rm max}\,[\text{Hz}]$ based on the central frequency of the $k$-th PRB is given as follows
\begin{align}
    f^\prime_{k,\rm max} = f_{c,\rm max} \left(1 + \frac{\Delta f_k}{f_c}\right) = 258 \left(1 + \frac{\Delta f_k}{28\,[\text{GHz}]}\right)
\end{align}
where $\Delta f_k$ is equal to the subtraction of the $k$-th PRB's central frequency from the carrier frequency. 
% In Fig.~\ref{fig:ch_resp}, we illustrate the frequency selectivity feature of the channel model and the dependency of the Doppler shift on PRBs' central frequencies. Fig.~\ref{fig:ch_resp}\,(a) depicts the mean normal amplitude and phase of channel statistics among $64$ PRBs at two sample slots of $100$ and $500$, while Fig.~\ref{fig:ch_resp}\,(b) shows their small-scale fading. 
% Owing to these figures, we can verify the model's frequency- and time-dependency properties.
% \begin{figure}[t!]   
% \centering
%     \subfloat[]{
%     \centering
%     \includegraphics[width=0.17\textwidth]{figures/channel_response.png}}
%     \hfil
%     \subfloat[]{  
%     \centering
%     \includegraphics[width=0.17\textwidth]{figures/small_scale_fading.png}}
%     \caption{Frequency-selective (a) channel statistics and corresponding (b) small-scale fading among $64$ PRBs at the $100$-th and $500$-th slots.}
%     \label{fig:ch_resp}
% \end{figure}

\subsection{Results and Discussion}
\begin{figure*}[t!]
    \centering
    \begin{subfigure}[t]{0.3\textwidth}
        \centering
        \includegraphics[width=1\textwidth, angle=0]{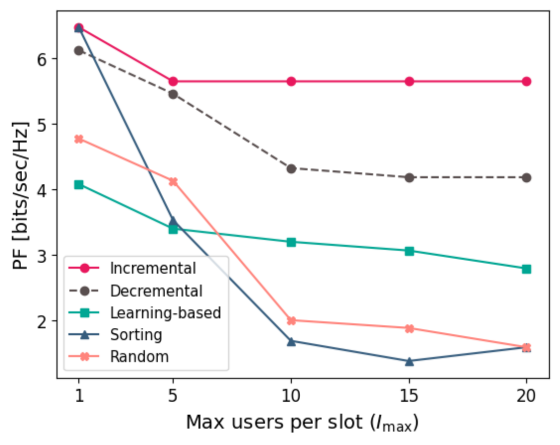}
        \caption{}
    \end{subfigure}
    \hfill
        \begin{subfigure}[t]{0.3\textwidth}
        \centering
        \includegraphics[width=1\textwidth, angle=0]{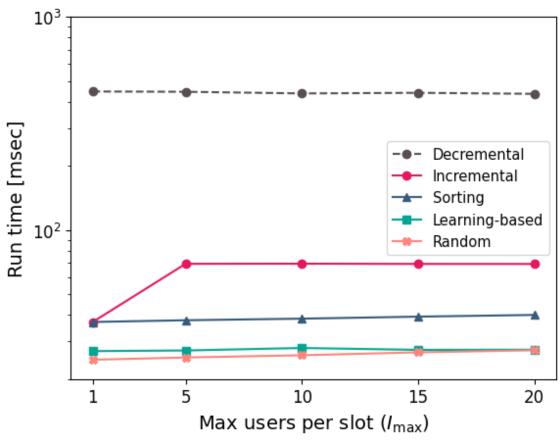}
        \caption{}
    \end{subfigure}
    \hfill
    \begin{subfigure}[t]{0.3\textwidth}
        \centering
        \includegraphics[width=1\textwidth, angle=0]{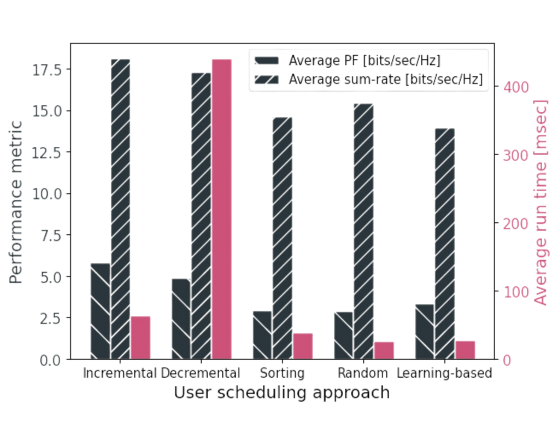}
        \caption{}
    \end{subfigure}
    \caption{The impact of the maximum number of users served per slot on (a) the average PF and (b) the required run time of the algorithms, alongside (c) the trade-off between their average performance and run time.}
    \label{fig:combined}
\end{figure*}

In Fig.~\ref{fig:combined}\,(a), we analyze the impact of increasing the maximum number of users that can be served per slot, i.e., $I_{\rm max}$, on the offered PF using different user scheduling approaches. The graph indicates that the incremental algorithm yields the highest PF, followed by the decremental algorithm. The learning-based approach shows moderate performance compared to the others, particularly for higher $I_{\rm max}$. Both greedy algorithms are flexible to refrain from selecting more users if adding them worsens the interference, leading to a reduced achievable rate and PF. The learning-based approach also possesses this adaptability. On average, the incremental, decremental, and learning-based approaches select $18.8\%$, $29.49\%$, and $16.62\%$ of the users, respectively. In contrast, the sorting and random approaches blindly schedule all possible users to be served. 
% \begin{figure}[t!]
%     \centering    
%     \includegraphics[width=0.33\textwidth, angle=0]{figures/PF_vs_max_users.png}
%     \caption{Interplay between the offered PF and the maximum number of users that can be served per slot.}
%     \label{fig:PF}
% \end{figure}
% \begin{figure*}
%     \centering
%     \begin{minipage}{0.31\textwidth}
%         \centering    
%         \includegraphics[width=1\textwidth, angle=0]{figures/PF_vs_max_users.png}
%         \caption{Interplay between PF and the maximum number of users served per slot.}
%     \end{minipage}
%     \hfill
%     \begin{minipage}{0.31\textwidth}
%         \centering
%         \includegraphics[width=1\textwidth, angle=0]{figures/runtime_vs_max_users.png}
%     \caption{Algorithms' run times versus the maximum number of users served per slot.}
%     \end{minipage}
%     \hfill
%     \begin{minipage}{0.31\textwidth}
%         \centering    
%         \includegraphics[width=1\textwidth, angle=0]{figures/PF_sumrate_runtime_single.png}
%         \caption{Trade-off between the offered performance and run time.}
%     \end{minipage}
% \end{figure*}

% The learning-based approach faces difficulty when $I_{\rm max}$ is small. Due to potential loss, the trained model initially selects a single user out of $20$, but its performance grows gradually for $I_{\rm max}\geq5$. Also, the reason that the tolerance of the offered PF by any approaches reduces for $I_{\rm max}>10$ is that when a large number of users are scheduled, the interference becomes strong enough that removing a user does not significantly affect the performance. Nevertheless, for $I_{\rm max}<10$, removing a user does make a difference in the presence of the noise power. Besides, scheduling all possible users, regardless of the interference they induce, leads to poor performance for the sorting and random approaches as $I_{\rm max}$ increases.
The run time of each user scheduling algorithm considering the maximum number of users that can be served is depicted in Fig.~\ref{fig:combined}\,(b). The decremental approach comes with a significant disadvantage in terms of long run time, resulting in much higher complexity than the others. The learning-based and random scheduling approaches require the shortest run time, independent of $I_{\rm max}$.
% since they do not schedule the users iteratively. 
The sorting algorithm takes longer, and the incremental one lies in between, with its run time increasing as $I_{\rm max}$ increases from $1$ to $5$. After that, the run time remains constant, as the algorithm chooses not to select more users to keep the induced interference low. The same reasoning applies to the decremental approach for $I_{\rm max}>10$.
% \begin{figure}[t!]
%     \centering    
%     \includegraphics[width=0.34\textwidth, angle=0]{figures/runtime_vs_max_users.png}
%     \caption{Run time required by different scheduling algorithms regarding the maximum number of users can be served per slot.}
%     \label{fig:run-time}
% \end{figure}

In Fig.~\ref{fig:combined}\,(c), the \emph{trade-off} between performance, measured as the average PF or sum-rate, and complexity, in terms of the average run time, for different user scheduling approaches is demonstrated. The most suitable approach is selected based on the key criterion of interest. For example, the incremental (decremental) approach provides $1.76$ ($1.47$) times higher PF at the cost of $2.3$ ($16.1$) times longer run time compared to the learning-based one. Therefore, greedy scheduling is selected \emph{if} a higher PF is more important; otherwise, learning-based scheduling might offer a much shorter run time.
% \begin{figure}[t!]
%     \centering    
%     \includegraphics[width=0.38\textwidth, angle=0]{figures/PF_sumrate_runtime_single.png}
%     \caption{Trade-off between the offered performance and run time.}
%     \label{fig:trade-off}
% \end{figure}

\begin{figure}[t!]
    \centering
    \includegraphics[width=0.3\textwidth, angle=0]{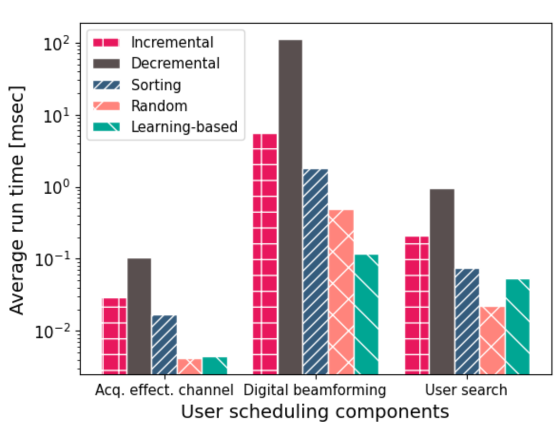}
    \vspace{-0.1cm}
    \caption{Timing profiles of the main user scheduling components.}
    \label{fig:algos_profile}
\end{figure}

Moreover, Fig.~\ref{fig:algos_profile} displays the timing profiles of the three main components involved in scheduling users within a short-length block. We observe that the longer time the incremental and decremental algorithms take to run is not only due to user search but also because of the time needed to obtain effective channel matrices and calculate the precoding matrix in each iteration. This confirms the higher complexity of the greedy incremental and decremental algorithms (see Section~\ref{sec:5-a}).

% In Fig.~\ref{fig:beam_allocate}, we assess how each user scheduling approach considers the \emph{beam angular separation} of the users for $I_{\rm max}=5$. The subfigures depict the azimuth and elevation angles of the analog beams assigned to the selected and deselected users for different scheduling approaches. It is shown that incremental and learning-based scheduling select users with large beam angular distances. These methods are advantageous because they support the angular separation of users, even when dealing with large $I_{\rm max}$. On the other hand, since the sorting algorithm schedules all users, as shown in Fig.~\ref{fig:beam_allocate}\,(c), the selected users have small beam angular distances, resulting in higher interference between those users. 
% % The same situation occurs with random scheduling.
% \begin{figure}[t!]   
% \centering
%     \subfloat[]{
%     \centering
%     \includegraphics[width=0.155\textwidth]{figures/beam_incremental.png}}
%     \hfill
%     % \subfloat[]{  
%     % \centering
%     % \includegraphics[width=0.11\textwidth]{figures/beam_decremental.png}}
%     % \hfill
%     \subfloat[]{
%     \centering
%     \includegraphics[width=0.155\textwidth]{figures/beam_sorting.png}}
%     \hfill
%     \subfloat[]{  
%     \centering
%     \includegraphics[width=0.155\textwidth]{figures/beam_learning.png}}
%     \caption{Angular separation of the selected and deselected users' analog beams after scheduling based on (a) incremental, (b) sorting, and (c) learning-based approaches for $I_{\rm max}=5$.}
%     \label{fig:beam_allocate}
% \end{figure}

\section{Conclusion}
We studied the multiuser scheduling challenge in mmWave MIMO-OFDM systems with hybrid beamforming. We defined PF as a long-term evaluation metric and formulated the optimization problem aimed at maximizing PF by optimally designing the digital and analog precoders, as well as scheduling the users to be served by the BS. To achieve this, we tailored a two-timescale protocol and implemented various scheduling methods, including greedy and learning-based approaches. Our results demonstrated the trade-off between performance and computational complexity for each approach. Additionally, we observed that the learning-based approach strikes a favorable balance between the average PF and the required run time. 

\balance
\bibliographystyle{IEEEtran}
\bibliography{references_EuCNC}

\end{document}